\documentclass[a4paper,aps,pre,superscriptaddress,floatfix,nofootinbib,twocolumn]{revtex4-1}

\usepackage{bbold}
\usepackage{bbm}
\usepackage[pdftex]{graphicx}
\usepackage{latexsym,amsmath,verbatim}
\usepackage[colorlinks,linkcolor=blue,citecolor=magenta,urlcolor=black]{hyperref}
\usepackage[svgnames]{xcolor}
\usepackage{rotating}
\usepackage{verbatim}
\usepackage{multirow}
\usepackage[english]{babel}
\usepackage{comment}
\usepackage[10pt]{moresize}
\usepackage{amsmath,amssymb}
\usepackage{pdfpages}

\makeatletter
\AtBeginDocument{\let\LS@rot\@undefined}
\makeatother

\begin{document}

\title{Optimal percolation on multiplex networks}

\author{Saeed Osat}
\affiliation{Molecular Simulation Laboratory, Department of Physics, Faculty of Basic Sciences, Azarbaijan Shahid Madani University, Tabriz 53714-161, Iran}

\author{Ali Faqeeh}
\affiliation{Center for Complex Networks and Systems Research, School
  of Informatics and Computing, Indiana University, Bloomington,
  Indiana 47408, USA}

\author{Filippo Radicchi}
\affiliation{Center for Complex Networks and Systems Research, School
  of Informatics and Computing, Indiana University, Bloomington,
  Indiana 47408, USA}
\email{filiradi@indiana.edu}

\begin{abstract}
Optimal percolation is the problem of finding the minimal set of nodes such that
if the members of this set are removed from a network, the network is fragmented
into non-extensive disconnected clusters. The solution of the optimal percolation problem
has direct applicability in strategies of immunization
in disease spreading processes, and influence maximization for certain classes of opinion dynamical models.
In this paper, we consider the problem of optimal percolation on multiplex networks.
The multiplex scenario serves to realistically model various technological, biological, and social networks.
We find that the multilayer nature of these systems, and more
precisely multiplex characteristics such as edge overlap and
interlayer degree-degree correlation, profoundly changes the
properties of the set of nodes
identified as the solution of the optimal percolation
problem.
\end{abstract}

\maketitle

\section{Introduction}

A multiplex is a network where nodes are connected through different
types or flavors of pairwise edges~\cite{boccaletti2014structure,kivela2014multilayer, lee2015towards}.~A convenient way to think of a multiplex is as a collection of network layers, each representing a specific type of edges.~Multiplex networks are genuine representations for
several real-world systems, including social~\cite{szell2010multirelational, mucha2010community}, and
technological systems~\cite{barthelemy2011spatial,cardillo2013emergence}.
From a theoretical point of view, a common strategy to understand the
role played by the co-existence of multiple network layers is based on a rather simple approach.~Given a process and a multiplex network, one studies the
process on the multiplex and on the single-layer projections of the
multiplex (e.g., each of the individual layers, or the network obtained from aggregation of the layers).
Recent research has demonstrated that
accounting for or forgetting about the effective
co-existence of different types of
interactions may lead to the emergence of rather
different features, and have potentially dramatic consequences
in the ability to model and predict properties of the system.~Examples include dynamical
processes, such as diffusion~\cite{PhysRevLett.110.028701,
  de2014navigability},
epidemic spreading~\cite{PhysRevE.85.066109, PhysRevE.86.026106,
  PhysRevLett.111.128701, de2016physics},
synchronization~\cite{delGenioe1601679}, and
controllability~\cite{PhysRevE.94.032316}, as well as
structural processes such as those typically framed
in terms of percolation models~\cite{buldyrev2010catastrophic,
  parshani2010interdependent, parshani2011inter, baxter2012avalanche,
  PhysRevE.89.012808, son2012percolation, min2014network,
  PhysRevE.91.012804, bianconi2014multiple, radicchi2013abrupt,
  radicchi2015percolation, cellai2016multiplex, PhysRevE.94.060301,
  radicchi2016redundant}.

The vast majority of the work on structural processes on
multiplex networks have focused on ordinary percolation models where
nodes (or edges) are considered either in a functional or in a non-functional state with homogenous probability~\cite{stauffer1991introduction}.
In this paper, we shift the focus on the optimal version
of the percolation process: we study the problem of
identifying the smallest number of nodes
in a multiplex network such that, if these nodes are removed,
the network is fragmented into many
disconnected clusters of non-extensive size. We refer to the nodes
belonging to this minimal set as Structural Nodes (SNs) of the multiplex network.
The solution of the optimal percolation problem has direct
applicability in the context of
robustness, representing the cheapest way to dismantle a network \cite{Mugisha16,Braunstein16,Zdeborova16}.
The solution of the problem of optimal percolation is, however,
important in other contexts, being equivalent to
the best strategy of immunization to a spreading process,
and also to the best strategy of seeding a network
for some class of opinion dynamical
models~\cite{Altarelli13,Clusella16,morone2015influence,pei2016collective}.
Despite its importance,
optimal percolation
has been introduced and considered
in the framework of single-layer networks
only recently~\cite{Clusella16,morone2015influence}.
The optimal percolation is an NP-complete problem~\cite{Braunstein16}. Hence, on large
networks, we can only use heuristic methods to find approximate
solutions.
Most of the research activity on this topic has indeed focused on the
development of greedy algorithms~\cite{Braunstein16, Clusella16,Mugisha16, Zdeborova16}.
The generalization of optimal percolation to multiplex networks that
we consider here consists in the
redefinition of the problem
in terms of mutual connectedness~\cite{buldyrev2010catastrophic}.
To this end, we reframe several algorithms for optimal percolation
from single-layer to multiplex networks.~Basically all the algorithms we use provide coherent solutions to
the problem, finding sets of SNs that are almost identical.
Our main focus, however, is not on the development of new algorithms, but
on answering the following question:
What are the consequences of neglecting the multiplex
nature of a network under an optimal percolation process?
We compare the actual solution of the optimal percolation problem
in a multiplex network with the solutions to the same problem
for single-layer networks extracted from the multiplex system.
We show that ``forgetting'' about the presence of multiple
layers can be potentially dangerous, leading to the overestimation
of the true robustness of the system mostly due to the
identification of a very high number of
false SNs. We reach this conclusion
with a systematic analysis of
both synthetic and real multiplex networks.

\section{Methods}
We consider a multiplex network composed of $N$ nodes
arranged in two layers. Each layer is an undirected and
unweighted network.~Connections of the two layers
are encoded in the adjacency matrices $A$ and $B$. The
generic element $A_{ij} = A_{ji}
=1$ if nodes $i$ and $j$ are connected in the first layer, whereas
 $A_{ij} = A_{ji}
=0$, otherwise. The same definition applies to the second layer, and
thus to the matrix $B$. The aggregated network
obtained from  the superposition of the two
layers is characterized by the adjacency matrix $C$,
with generic
elements $C_{ij} = A_{ij} + B_{ij} - A_{ij} B_{ij}$.
The basic objects we look at are clusters
of mutually connected nodes~\cite{buldyrev2010catastrophic}:
Two nodes in a multiplex network are mutually connected, and
thus part of the same cluster of mutually connected nodes, only
if they are connected by at least a path, composed of
nodes within the same cluster, in every layer of  the system.
In particular, we focus our attention on the largest among these
cluster,
usually referred to as the Giant Mutually Connected Cluster (GMCC).
Our goal is to find the minimal set of nodes that, if removed from the
multiplex,
leads to a GMCC that has at maximum a size equal to $\sqrt{N}$.
This is a common prescription, yet not the only one possible, to
ensure that all clusters have non-extensive sizes in systems with a
finite number of elements~\cite{Clusella16}.
Whenever we consider single-layer networks, the above
prescription apply to the single-layer clusters in the same exact way.

\begin{figure}\centering
\includegraphics[width=0.29\textwidth]{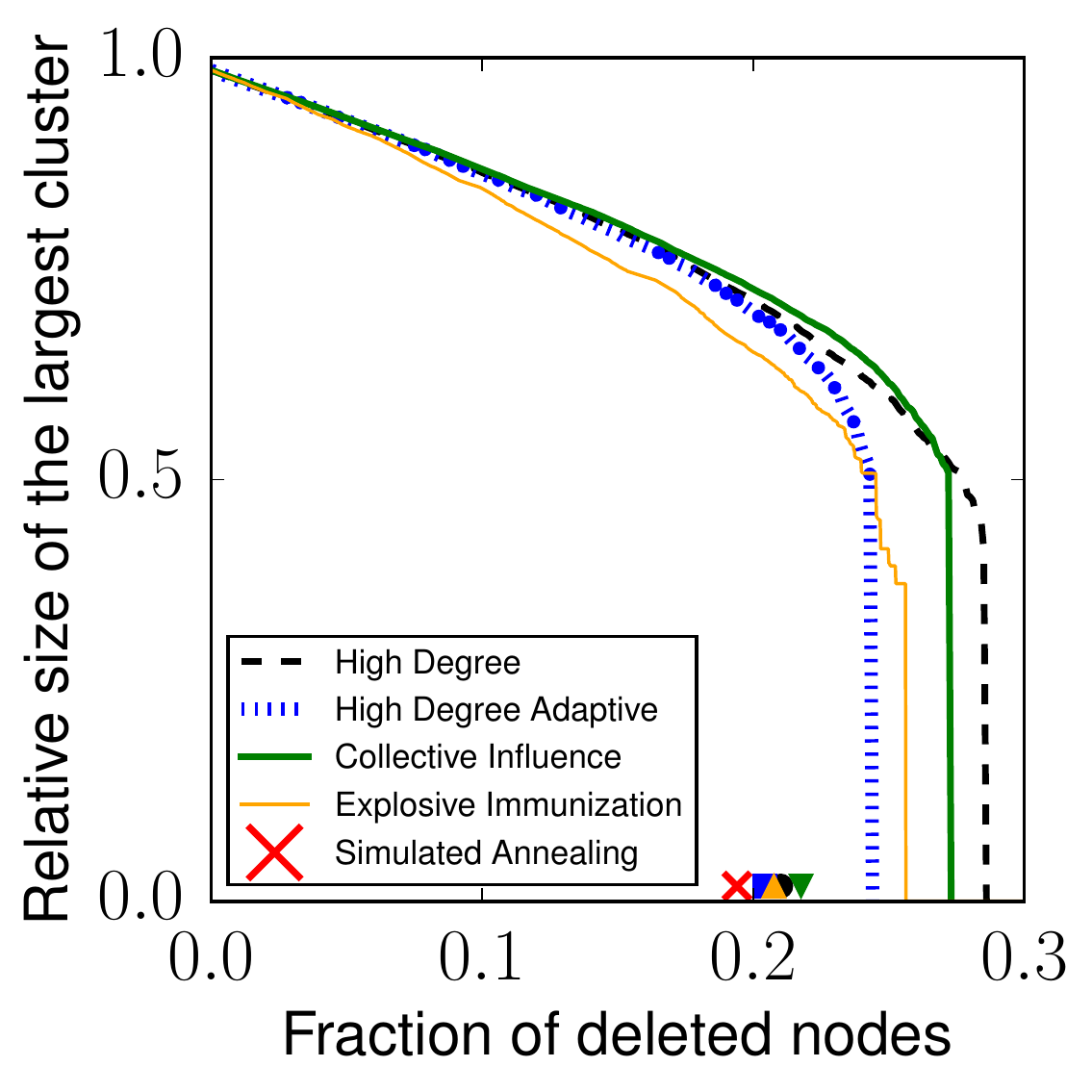}
\caption{Comparison among
different algorithms to approximate solutions
of the optimal percolation problem. We consider a multiplex
network with $N=10,000$ nodes. The multiplex is composed of two
network layers generated independently according to
the Erd\H{o}s-R\'enyi model with
average
degree $\langle k \rangle = 5$. Each curve
represents the relative size of the GMCC
as a function of the number of nodes
inserted in the set of SNs, thus removed from the
multiplex. Colored markers indicate the effective fraction of nodes
left in the set of SNs after a greedy post-processing
technique is applied to the set found by the corresponding algorithm.
The red cross identifies instead the size of the set of SNs found trough the Simulated Annealing optimization.
Please note that the ordinate value of the markers
has no meaning; in all cases, the relative size of the
largest cluster is smaller than $\sqrt{N}$.
Details on the implementation of the
various algorithms are provided in the SI.}
\label{fig1}
\end{figure}

We generalize most of the algorithms devised to
find approximate solutions to the optimal percolation
problem in
single-layer networks to
multiplex networks~\cite{morone2015influence, Braunstein16,
Clusella16,
  Mugisha16, Zdeborova16}. Details on the
implementation of the various methods are provided in the
Supplementary Information (SI).
We stress that the generalization of these methods
is not trivial at all.
For instance, most of the greedy methods use node degrees
as crucial ingredients. In a multiplex network, however,
a node has multiple degree values, one for every layer.
In this respect, it is not clear
what is the most effective way of combining these numbers
to assign a single score to a node: they may be summed, thus
obtaining a number approximately equal to the degree of the node in
the aggregated network derived from the multiplex, but also multiplied, or combined in more complicated ways.
We find that the results of the
various algorithms are not particularly sensible to this choice,
provided that a simple post-processing technique is applied to the set of
SNs found by a given method.
In Figure~\ref{fig1} for example, we show
the performance of several greedy algorithms when applied to a
multiplex network composed of two layers
generated independently according to the Erd\H{o}s-R\'enyi (ER) model.
Although the mere application of an algorithm may lead to
different estimates of the size of the set of SNs, if we
greedily remove from these sets the nodes that do not increase
the size of the GMCC to the predefined sub-linear threshold ($\sqrt{N}$)~\cite{Zdeborova16}, the sets obtained after
this post-processing technique have almost identical sizes.

As Figure~\ref{fig1} clearly shows, the best results, in the sense that the
size of the set of SNs is minimal, is found with a Simulated
Annealing (SA) optimization strategy~\cite{Braunstein16} (see details in the SI).
The fact that the SA method is outperforming score-based algorithms
is not surprising. SA actually represents one of the
best strategies that one can apply in hard optimization tasks.
In our case, it provides us with a reasonable upper-bound of the size
of the set of SNs that can be identified in a multiplex network.
The second advantage of SA in our context is that it doesn't rely on
ambiguous definitions of ingredients (as for example, the
aforementioned issue of node degree).
Despite its better performance, SA has a serious
drawback in terms of computational speed.
As a matter of fact, the algorithm can be applied only
to multiplex networks of moderate size. As here we are interested in
understanding
the fundamental properties of the optimal percolation problem in
multiplex networks, the analysis presented in the main text of the
paper is entirely based on results obtained
through SA optimization. This provides us with a solid ground to
support our statements. Extensions, relying on score-based algorithms, of the same analyses to larger multiplex networks are qualitatively similar (see SI).

\section{Results}

\begin{figure}
\includegraphics[width=0.5\textwidth]{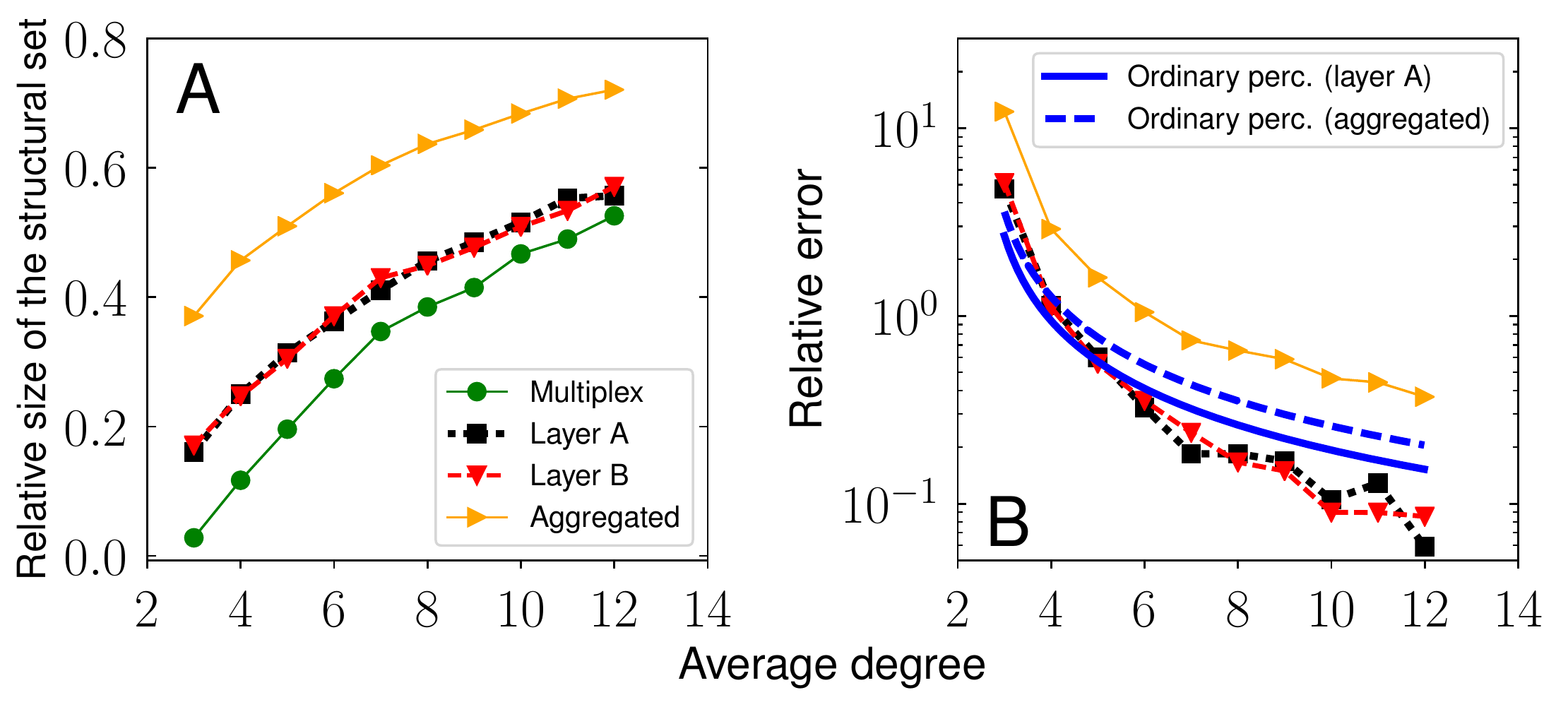}
\caption{Optimal percolation problem in synthetic multiplex networks.~A) We consider multiplex networks
with $N=1,000$ and layers generated independently
according to the Erd\H{o}s-R\'enyi model with average degree $\langle
k \rangle$.~We estimate the relative size of the set of SNs on the multiplex as a function of $\langle k \rangle$ (green
circles), and compare with the same quantity but estimated on the
individual layers (black squares, red down triangles) or the aggregated
(orange right triangles). B) Relative errors of
single-layer estimates of the size of the structural set
with respect to the ground-truth value provided by the
multiplex estimate. Colors and symbols are the same as those used
in panel A. The blue curves with no markers represent instead the results for an ordinary site percolation process~\cite{buldyrev2010catastrophic}.
}
\label{fig2}
\end{figure}

\subsection{The size of the set of structural nodes}
We consider the relative size of the
set of SNs, denoted by $q$, for a multiplex composed
of two independently fabricated ER network layers as a function of their
average degree $\langle k \rangle$.~We compare the results obtained
applying the SA algorithm to the multiplex, namely $q_M$, with those obtained
using SA on the individual layers, i.e., $q_A$ and $q_B$, or the aggregated
network generated from the superposition of the two layers,
i.e., $q_S$. By definition we expect that $q_M \leq  q_A
\simeq q_B \leq q_S$.  What we don't know, however, is how bad/good are the measures $q_A$, $q_B$ and $q_S$ in the prediction of the effective
robustness of the multiplex $q_M$. For ordinary random percolation
on ER multiplex networks with negligible overlap, we know that $q_M \simeq 1- 2.4554 / \langle k\rangle$~\cite{buldyrev2010catastrophic}, $q_A \simeq q_B \simeq 1- 1 /\langle k\rangle$, and $q_S \simeq 1-1/(2\langle k\rangle)$~\cite{molloy1995critical}.
Relative errors are therefore $\epsilon_A \simeq \epsilon_B \simeq (2.4554 - 1) / (\langle k \rangle -2.4554)$, and $\epsilon_S \simeq (2.4554 - 1/2) / (\langle k \rangle -2.4554)$.~We find that the relative error for the optimal percolation behaves more or less in the same way as that of the ordinary percolation (Figure~\ref{fig2}B), noting that, as $\langle k \rangle$ is increased, the decrease in the relative error associated with the individual layers is slightly faster than what expected for the ordinary percolation.~The relative error associated with the aggregated network is instead larger than the one expected from the theory of ordinary percolation.~As shown in Figure~\ref{fig2}A, for sufficiently large $\langle k\rangle$, dismantling the ER multiplex network is almost as hard as dismantling any of its constituent layers.

\begin{figure}
\includegraphics[width=0.3\textwidth]{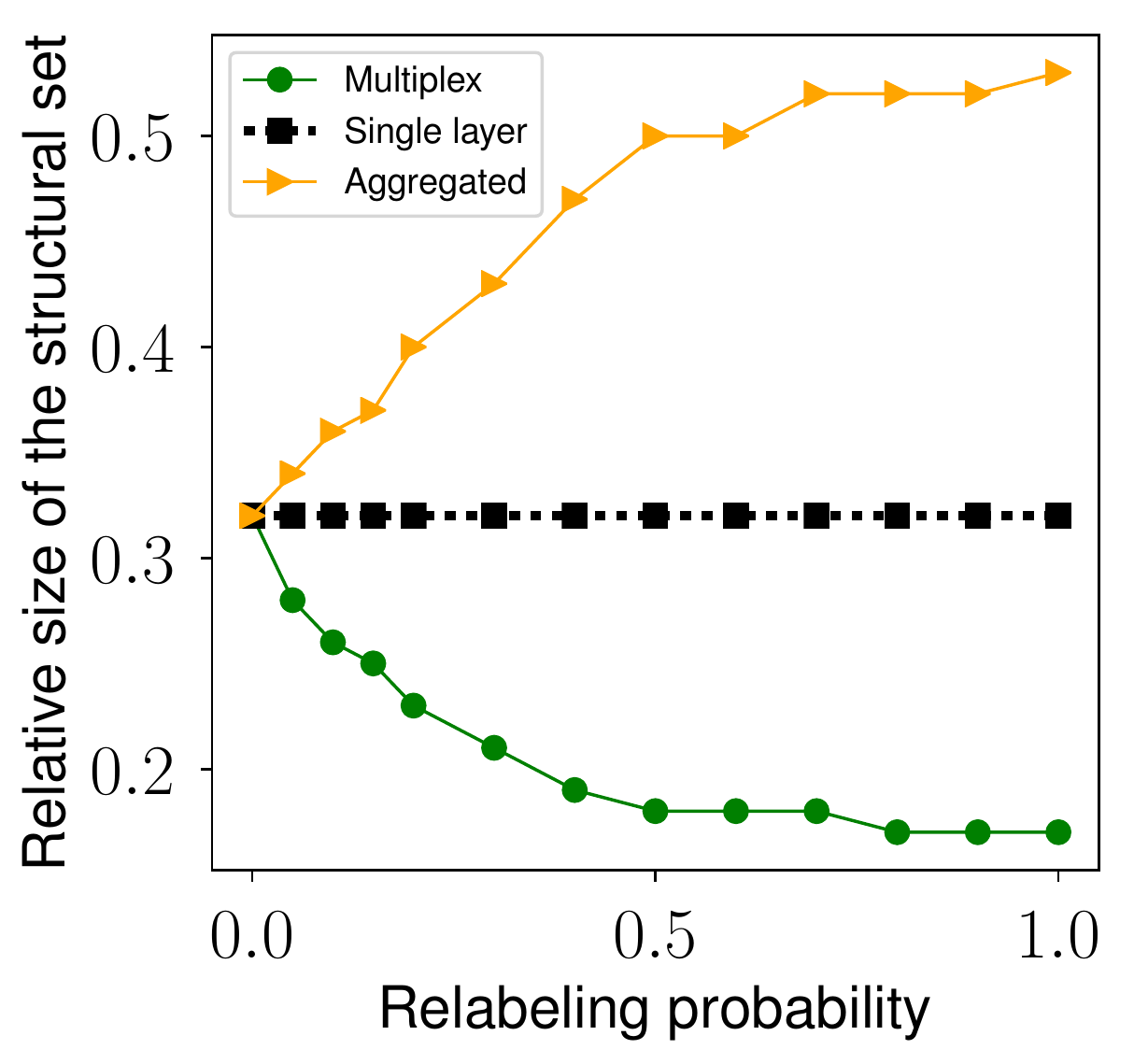}
\caption{The effect of reducing edge overlaps and interlayer degree-degree correlation by partially relabeling nodes in multiplex networks with initially identical layers.
Initially, both layers are a copy of a random network generated by an Erd\H{o}s-R\'enyi model with $N=1,000$ nodes and average degree $\langle k \rangle=5$.
Then, in one of the layers, each node is selected to switch its label with another randomly chosen node with a certain probability $\alpha$. For each $\alpha$, we determine the mean of the relative size of the set of SNs over 100 realizations of the SA algorithm on the multiplex network.} \label{fig3}
\end{figure}

\subsection{Edge overlap and degree correlations}
Next, we test the role played by edge overlap and layer-to-layer
degree correlation in the
optimal percolation problem.~These are ingredients
that dramatically change the nature of the
ordinary percolation transition
in multiplex networks~\cite{cellai2013percolation, bianconi2013statistical, min2015link,  radicchi2015percolation,
  baxter2016correlated, PhysRevE.94.032301}.
 In Figure~\ref{fig3}, we report
results of a simple analysis.~We take advantage of the
model introduced in Ref.~\cite{bianconi2014percolation}, where a multiplex
is constructed with two identical layers.~Nodes in one of the two
layers relabeled with a certain probability $\alpha$.
For $\alpha=0$, multiplex, aggregated network and single-layer
graphs are all identical. For $\alpha=1$, the
networks are analogous to those considered in the previous section.
We note that this model doesn't
allow to disentangle the role played by edge overlap among layers
and the one played
by the correlation of node degrees. For $\alpha = 0$, edge overlap
amounts to $100\%$, and there is a one-to-one match between the
degree of a node in one layer and the other. As $\alpha$ increases, both
edge overlap and degree correlation decrease simultaneously.
As it is apparent from the results of Figure~\ref{fig3}, the
system reaches the multiplex regime for very small values of $\alpha$,
in the sense that the relative size of the set of SNs deviates instantly from its value for $\alpha=0$.
This is in line with what already found in the context of ordinary percolation
processes in multiplex networks: as soon as there is
a finite fraction of edges that are not shared by the two layers, the
system behaves exactly as a
multiplex~\cite{cellai2013percolation, bianconi2013statistical, min2015link,  radicchi2015percolation,
  baxter2016correlated, PhysRevE.94.032301}.

\subsection{Accuracy and sensitivity}
\begin{figure}[b]
\includegraphics[width=0.5\textwidth]{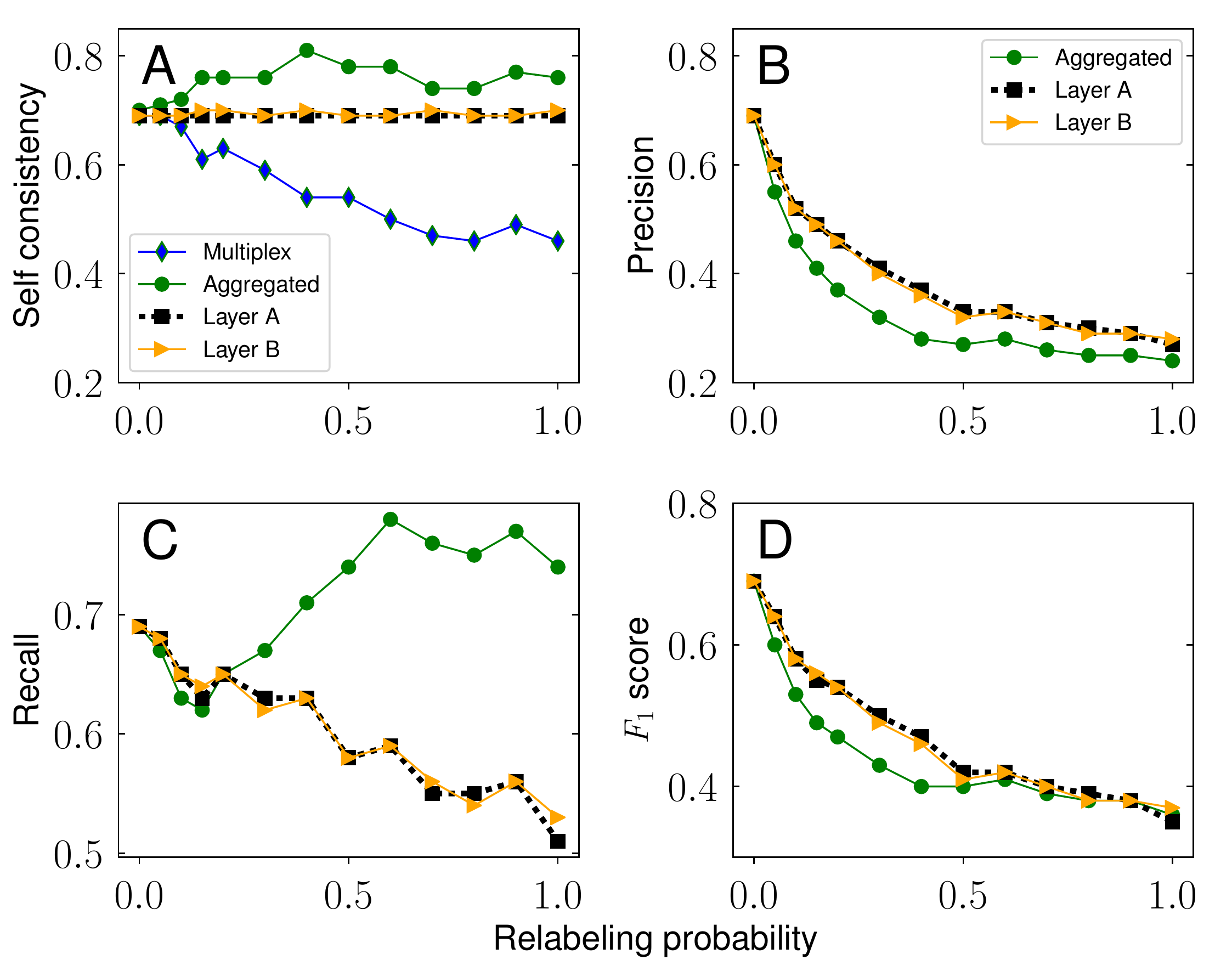}
\caption{The effect of reducing edge overlaps and interlayer degree-degree correlation by partially relabeling nodes in multiplex networks with initially identical layers.
We consider the multiplex networks described in Figure~\ref{fig2} and the sets of SNs found for the multiplex and single layer based representations of these networks.
A) As the set of SNs found in different instances of the optimization algorithm are different from each other, we first quantify the self-consistency of those solutions across $100$ independent runs of the SA algorithm.
We then assume that the multiplex representation provides the ground-truth classification of the nodes.~We compare the results of the other representation with the ground truth by measuring their precision (panel B), their sensitivity or recall (panel C), and their $F_{1}$ score (panel D).}
\label{fig4}
\end{figure}

So far, we focused our attention only on the size of the
set of SNs.~We neglected, however, any analysis
regarding the identity of the nodes that actually
compose this set. To proceed with such an analysis,
we note that
different runs of the SA algorithm (or any
algorithm with stochastic features) generally produce
slightly different sets of SNs (even if they all have
almost identical sizes).~The issue is not related to the
optimization technique, rather to the existence of
degenerate solutions to the problem. In this respect, we work with the quantities $p_i$, each of which describes the probability that a node $i$ appears in the set of SNs in a realization of the detection method (here, the SA algorithm).~This treatment takes into account the fact that a node may belong to the set of SNs in a number of realizations of the detection method and may be absent from this set in some other realizations.

Now, we define the self-consistency of a detection method as $S = \sum_{i} p_i^2 / \sum_{i} p_i$, which describes the ratio of the expected overlap between two SNs obtained from two independent realizations of the detection method to the expected size of an SN.
If the set of SNs is identical across different runs, then  $S=1$.~On the other hand, the minimal value we can observe is $S = Q/N$, assuming that the size of the structural set is equal to $Q$ in all runs, but nodes belonging to this set are changing all the times, so that for
every node we have $p_i=Q/N$.

As reported in Figure~\ref{fig4}A, even for random multiplex networks, self-consistency is rather high for single layer representations of the network.~On the other hand, $S$ decreases significantly as the overlap and interlayer degree correlations are  decreased (Figure~\ref{fig4}A). The low $S$ values for multiplexes with small overlap and correlation together with the small sizes of their set of SNs (Figure~\ref{fig2}) suggests that in such networks there can exist many relatively different sets of SNs that if nodes of each of these sets are removed the network is dismantled.

Next, we turn our attention to quantifying how the sets of SNs identified in single-layer or aggregated networks are representative for the ground-truth sets found on the multiplex networks.~Here, we denote by $p_i$ and $w_i$ the probability that node $i$ is found within the set of SNs of, respectively, a multiplex network (ground truth) and a specific single-layer representation of that multiplex.~To compare the sets represented by $w_i$ to the ground truth sets, we
adopt three standard metrics in information retrieval \cite{Chu05data_mining,Maino11_ICIAP}, namely precision, recall and the Van Rijsbergen's $F_1$ score:~Precision is defined as $P = [ \sum_{i} p_i w_i ] / [\sum_i w_i] $, i.e., the ratio of the (expected) number of correctly detected SNs to the (expected) total number of detected SNs.~Recall is instead defined as $R = [ \sum_{i} p_i w_i ] / [\sum_i p_i]$, i.e., the ratio of the (expected) number of correctly detected SNs to the (expected) number of actual SNs of the multiplex.~We note that the self-consistency we previously defined corresponds to precision and recall of the ground-truth set with respect to itself, thus providing a base line for the interpretation of the results.~The $F_1$ score defined as $F_1=\frac{2}{1/P+1/R}$ provides a balanced measure in terms of $P$ and $R$.

As Figure~\ref{fig4}B shows, precision deteriorates as the edge overlap and interlayer degree correlation decrease by increasing the relabeling probability.~In particular, when the overlap and correlation between the layers of the multiplex network are not large, the precision of the sets of SNs identified in single layers or in the superposition of the layers is quite small (around $0.3$), even smaller than the ratio of the $q_M$ of the multiplex to the $q$ of any of these sets (see Figure~\ref{fig3}).~This means that, when the multiplex nature of the system is neglected, not only too many SNs are identified, but also a significant number of the SNs of the multiplex are not identified.

Recall, on the other hand, behaves differently for single-layer and aggregated networks (Figure~\ref{fig4}C). In single layers, we see that recall systematically decreases as the
relabelling probability increases.~The structural set of nodes obtained on the superposition of the layers instead provides large values of recall. This is not due to good performance rather to the fact that
the set of SNs identified on the aggregated network is very large (see Figure~\ref{fig3}). The results of Figure~\ref{fig4} demonstrate that even the larger recall values for the aggregated network do not lead to a better $F_1$ score: the $F_1$ score diminishes as the relabeling probability is increased.

\begin{table*}
\footnotesize
\begin{tabular}{l | l | c | c c | c c c c | c c c c | c c c c}
\multirow{3}{*}{Network} & \multirow{3}{*}{~Layers} & \multirow{3}{*}{$N$} & \multicolumn{2}{|c|}{Multiplex} & \multicolumn{8}{|c}{Single layers~~~~} &  \multicolumn{4}{|c}{Aggregate}
\\
\cline{4-17}
\cline{4-17}
\multirow{2}{*}{~} & \multirow{2}{*}{~} & \multirow{2}{*}{~} & \multirow{2}{*}{$q_M$} & \multirow{2}{*}{$S$} & \multirow{2}{*}{$q_A$} & \multirow{2}{*}{$P_A$} & \multirow{2}{*}{$R_A$} & \multirow{2}{*}{$F_1^{(A)}$} & \multirow{2}{*}{$q_B$} & \multirow{2}{*}{$P_B$} & \multirow{2}{*}{$R_B$} & \multirow{2}{*}{$F_1^{(B)}$} & \multirow{2}{*}{$q_S$} & \multirow{2}{*}{$P_S$} & \multirow{2}{*}{$R_S$} & \multirow{2}{*}{$F_1^{(S)}$}\\
& & & & & & & & & & & &  &
\\
\hline
\hline
\multirow{3}{*}{\it Air Transportation}~\cite{radicchi2015percolation}  &
{\it American Air. -- Delta}&
 $84$ & $ 0.12 $ & $ 0.85 $ & $ 0.14 $ & $ 0.58 $ & $ 0.70 $ & 0.63 & $ 0.32 $ & $0.29$ & $0.79$ & 0.42 & $ 0.35 $ & $ 0.32 $ & $ 0.92 $ & 0.47
\\
&
{\it American Air. -- United}&
   $73$  &  $ 0.10 $     & $ 0.99 $     &  $ 0.16 $    &   $ 0.32 $    &      $ 0.52 $   & 0.40     &  $ 0.14 $       &   $ 0.68 $     &  $ 1.00$       & 0.81       &    $0.25$     &  $0.39$       & $1.00$ & 0.56
\\
&
{\it United -- Delta}&
  $82$   &   $0.10$     &   $1.00$     &   $0.27$    &   $0.23$    &     $0.62$    & 0.34     &   $0.12$      &   $0.80$     &   $1.00$   & 0.89         &  $0.33$       &   $0.30$      & $1.00$  & 0.46
\\

\hline
\multirow{3}{*}{\it C. Elegance}~\cite{chen2006wiring, de2014muxviz}  &
{\it Electric} -- {\it Chem.~Mon.} &
238      &    0.09     &     0.69            &   0.16    &   0.41    &   0.71    & 0.52      &    0.26     &     0.22    &   0.60  & 0.32   &    0.35     &     0.21    &   0.79  & 0.33
\\
&
{\it Electric} -- {\it Chem.~Pol.} &
252      &    0.12     &     0.79            &   0.15    &   0.50    &   0.63     & 0.56       &    0.39     &     0.24    &   0.78    & 0.37  &    0.45     &     0.22    &   0.82  & 0.35
\\
&
{\it Chem.~Mon.} -- {\it Chem.~Pol.} &
259      &    0.25     &     0.82            &   0.28    &   0.69    &   0.77    & 0.73        &    0.39     &     0.51    &   0.79     & 0.62       &    0.42     &     0.48    &   0.80 & 0.60
\\

\hline
\multirow{3}{*}{\it Arxiv}~\cite{de2015identifying}  &
{\it {\ssmall physics.data-an}} -- {\it  {\ssmall cond-mat.dis-nn}} &
1400      &    0.05     &     0.78            &   0.10    &   0.38    &   0.77      & 0.51     &    0.07     &     0.55    &   0.75     & 0.63       &    0.13     &     0.31    &   0.81 & 0.45
\\
&
{\it  {\ssmall physics.data-an}} -- {\it  {\ssmall cond-mat.stat-mech}} &
709       &    0.03     &     0.73           &   0.08    &   0.23    &   0.67    & 0.34       &    0.03     &     0.64    &   0.72     & 0.68      &    0.09     &     0.22    &   0.74 & 0.34
\\
&
{\it  {\ssmall cond-mat.dis-nn}} -- {\it  {\ssmall cond-mat.stat-mech}} &
499       &    0.02     &     0.50            &   0.06    &   0.13    &   0.46     & 0.20     &    0.04     &     0.23    &   0.51      & 0.32      &    0.09     &     0.13    &   0.65 & 0.22
\\

\hline
\multirow{3}{*}{\it Drosophila M.}~\cite{stark2006biogrid, de2015structural}  &
{\it Direct} -- {\it Supp.~Gen.} &
676      &    0.01     &     0.62            &   0.07    &   0.12    &   0.60  & 0.20         &    0.11     &     0.09    &   0.64 & 0.16         &    0.19     &     0.07    &   0.87 & 0.13
\\
&
{\it Direct} -- {\it Add.~Gen.} &
626      &    0.01     &     0.81            &   0.07    &   0.06    &   0.64  & 0.11         &    0.09     &     0.05    &   0.59  & 0.09          &    0.16     &     0.04    &   0.87 & 0.08
\\
&
{\it Supp.~Gen.} -- {\it Add.~Gen.} &
557      &    0.09     &     0.82            &   0.14    &   0.44    &   0.74   & 0.55       &    0.12     &     0.50    &   0.70  & 0.58         &    0.20     &     0.35    &   0.80 & 0.49
\\

\hline
\multirow{2}{*}{\it Homo S.}~\cite{de2014muxviz, stark2006biogrid}  &
{\it Direct} -- {\it Supp.~Gen.} &
4465     &    0.05     &     0.72            &   0.16    &   0.20    &   0.73 & 0.31          &    0.13     &     0.23    &   0.64   & 0.34         &    0.27     &     0.15    &   0.89 & 0.26
\\
&
{\it Physical} -- {\it Supp.~Gen.} &
5202     &    0.05     &     0.75            &   0.15    &   0.23    &   0.77 & 0.35           &    0.13     &     0.22    &   0.63 & 0.33          &    0.26     &     0.16    &   0.90 & 0.27
\\

\hline \hline
\end{tabular}
\caption{Optimal percolation in real multiplex networks.
From left to right we report the following information. The first three columns contain the name of the system, the identity of the layers, and the number of nodes of the network. The fourth and fifth columns are results obtained from the optimal percolation problem studied on the multiplex network, and contain information about the relative size $q_M$, and self-consistency metric $S$ of the set of SNs.
Then, we report results obtained for the first single-layer network of
the multiplex, namely the fraction $q_A$ of nodes in the structural set, the precision $P_A$, the recall $R_A$, and the $F_1$ score of the set of SNs of the first layer. The next three columns are identical to those, but refer to the second layer. Finally, the three rightmost columns contain information about the fraction $q_S$ of nodes in the structural set, $P_S$ precision, $R_S$ recall, and the $F_1$ score of the set of SNs for the aggregated network obtained from the superposition of the two layers. All results have been obtained with $100$ independent instances of the SA optimization algorithm.} \label{tab1}
\end{table*}

\subsection{Real-world multiplex networks}
\begin{figure}[hb!]
\includegraphics[width=0.45\textwidth]{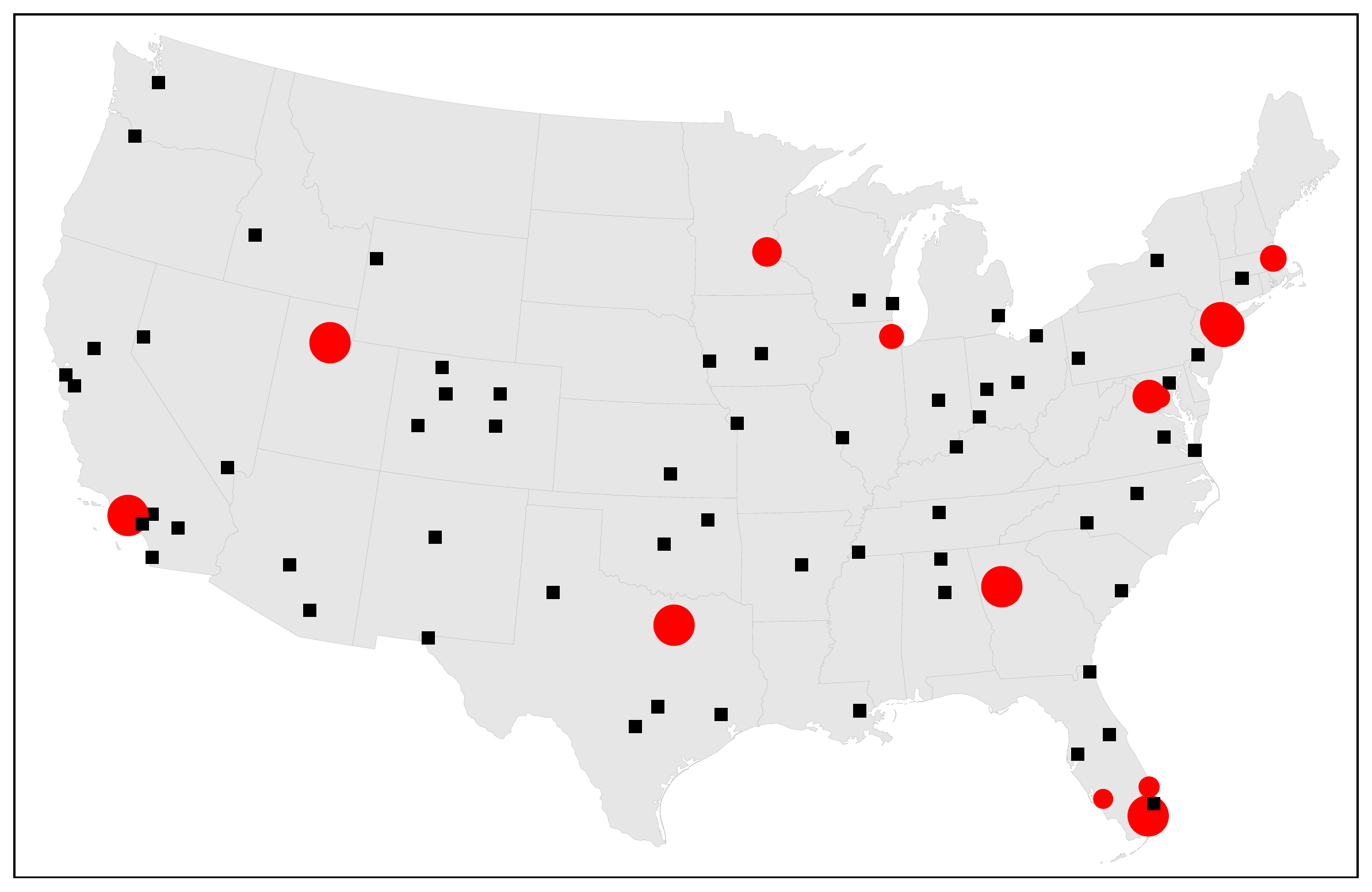}
\caption{Optimal percolation on the multiplex network of US domestic flights operated in January 2014 by {\it American Airlines} and {\it Delta}.
The red circles represent the nodes that were a member of the set of
structural nodes in different realizations of the optimal percolation
on the multiplex representation of the network.~The size of each
circle is proportional to the probability of finding that node in the
set of SNs.~All other airports in the multiplex are represented as
black squares. Interestingly, not all the 14 structural nodes match
the top 14 busiest {\it hubs} \cite{webpage:airports}, nor the
probabilities follow the same order as the flight traffic of these  airports.~The results have been obtained with $100$ independent instances of the SA optimization algorithm.} \label{fig5}
\end{figure}
In Table~\ref{tab1}, we present results of the analysis of optimal
percolation
problem on several real-world multiplex networks generated from empirical data.~For most of these networks, the optimal percolation on the multiplex
representation has a rather high self-consistency. This implies that
there is a certain small group of nodes that have a major importance
in the robustness of such real-world networks to the optimal
percolation process. The $F_1$ score for most of the networks (not
shown) is quite low indicating that on real-world networks we loose
essential information about the optimal percolation problem if the
multiplex structure is not taken into account.

To provide a practical case study with a lucid interpretation, we depict, in Figure~\ref{fig5},  the results for optimal percolation on
a multiplex network describing
the air transportation operated by two of the major airlines in the
United States. SA identifies always $10$ airports in the set of
SNs. There is a slight variability among different instances of the
SA optimization, with a total of $14$ distinct airports appearing
in the structural set at least once over $100$ SA instances. However,
changes in the SN set from run to run mostly regard
airports in the same geographical region.~Overall, airports in the
structural set are scattered homogeneously across the country, suggesting that the GMCC of the network mostly relies on hubs serving specific
geographical regions, rather than global hubs in the entire
transportation system. For instance, the probabilities
that describe the membership of the airports to the set of SNs do not
strictly follow the same order as that of the recorded flight traffics~\cite{webpage:airports}; nor merely the number of connections of the airports (not shown) is sufficient to determine the structural
nodes.~This is well consistent with the collective nature of the optimal percolation on the complex network of air transportation.

\section{Conclusions}
In this paper, we studied the optimal percolation problem on
multiplex networks. The problem regards the detection of
the minimal set of nodes (or set of structural nodes, SNs)
such that if its members are removed from the network,
the network is dismantled. The solution to the problem
provides important information on the microscopic parts
that should be maintained in a functional state
to keep the overall system functioning,
in a scenario of maximal stress.
Our study focused mostly on the characterization of the SN sets
of a given multiplex network in comparison with those found on the
single-layer projections of the same multiplex, i.e., in a scenario
where one ``forgets''  about the multiplex nature of
the system. Our results demonstrate that, generally, multiplex
networks have considerably
smaller sets of SNs
compared to the SN sets of their single-layer based network
representations. The error committed when relying on
single-layer representations of the multiplex doesn't regard only the size of the
SN sets, but also the identity of the SNs.
Both issues emerge in the
analysis of synthetic network models, where
edge overlap and/or interlayer degree-degree
correlations seem to fully explain the
amount of discrepancy between the SN set of a multiplex
and the SN sets of its single-layer based representations.
The issues are apparent also in many of the real-world
multiplex networks we analyzed. Overall,  we conclude that neglecting
the multiplex structure of a
network system subjected to maximal structural stress
may result in significant inaccuracies about its
robustness.

\begin{acknowledgements}
AF and FR acknowledge support from the US Army Research Office (W911NF-16-1-0104).
FR acknowledges support from the National Science Foundation (Grant
CMMI-1552487). 
\end{acknowledgements}


%

\clearpage
\newpage


\section*{Supplementary material}

# Optimal percolation on multiplex networks

Saeed Osat, Ali Faqeeh, Filippo Radicchi

# Supplementary Information

## CONTENTS

## S.1. DISMANTLING ALGORITHMS

In this section, we briefly discuss some of the most effective dismantling algorithms on monoplex networks and their generalization to multiplex networks with two layers. The aim of these algorithms is to approximate the set of structural nodes which is the minimal set of nodes that their removal dismantles the network into vanishingly small (non-extensive) clusters. We first introduce some of the score-based algorithms. In such algorithms, at each step, a score for each node is calculated and the node with the highest score is removed (when several nodes have the same score, one of them is removed at random). We discuss four different types of such algorithms: High Degree (HD), High Degree Adaptive (HDA), Collective Influence (CI) and Explosive Immunization (EI). These methods are partially deterministic in nature, i.e., nearly the same set of structural nodes are discovered at each realizations of the algorithm. Besides score-based algorithms, we present Simulated Annealing (SA) algorithm which is a greedy algorithm that searches the solution space of the dismantling problem to find the best approximation to the structural sets. The SA method takes into account the collective behavior of the dismantling problem and provides several dismantling sets for different realizations of the algorithm on the same network structure.

## A.  High Degree (HD)

In a monoplex network, the easiest way to dismantle a network, is a degree-based attack. After sorting the nodes with respect to their degrees, the nodes with the highest degree are removed one by one[1] until the network is dismantled. As this algorithm is deterministic (except from the randomness in choosing a node from those with the same degree), the set of nodes that are removed to dismantle the network is almost unique.

In multiplex networks, the degree of a node can be defined in various ways. We consider four different cases: the score of a node is defined as (i) its degree in layer A, (ii) its degree in layer B, (iii) the sum of its degrees across all the layers, and (iv) the product of its degrees across all the layers. It is worth mentioning that, when using HD, HDA and CI methods, at each step we remove $0.001 \times N$ of the nodes, where $N$ is the total number of nodes (in

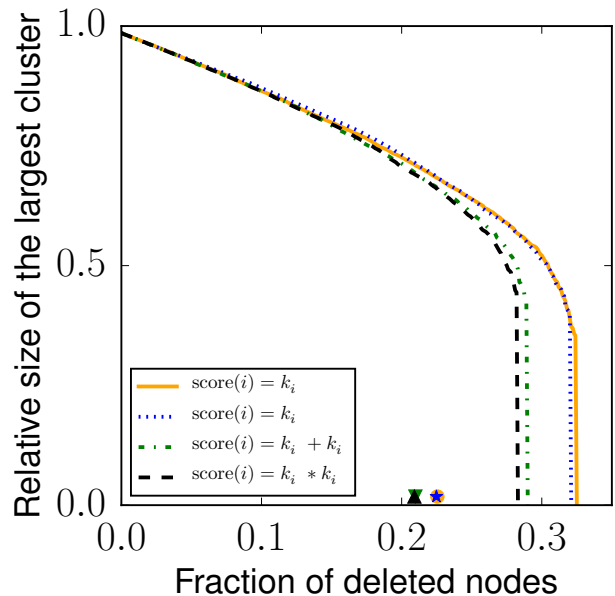

Figure S-1. Optimal percolation for a multiplex network using the High Degree (HD) algorithm. We consider a multiplex network composed of two identical layers with $N = 10000$ nodes generated according to an Erdős–Rényi model with average degree $\langle k \rangle = 5.0$. Different line styles correspond to four different methods of defining node scores in multiplex networks. Markers are the dismantling fraction obtained with these four methods when combined with the Greedy Reinserting (GR) procedure (see Sec. S.2).

---

[1] In cases where there are more than one node with a certain degree, one of them is removed at random.

each layer) of the multiplex.

As Figure S-1 illustrates, to destruct a multiplex, the two scores defined as a combination of degrees in different layers are more effective than those based on the degrees in only one of the layers. In the main script and in the rest of the Supplemental Material (SM) when we refer to HD method, we mean the one in which a node degree is the product of its degrees across all the layers.

### B.  High Degree Adaptive (HDA)

In the HD algorithm if we take into account the history of the process and recalculate, at each step, the degrees of the nodes, it is referred to as an HDA algorithm. Since the HDA algorithm is adaptive it is expected to work better than the HD method. In each step of the monoplex version of the HDA, we remove a fraction $0.001 \times N$ of the nodes that had the highest degrees; then we recalculate the degree of the nodes present in the Giant Connected Component (GCC) of the network. We repeat this process until the size of the GCC reduces to $\sqrt{N}$ or smaller; this threshold satisfies the condition that in the dismantled network the size of all the clusters is a sub-linear function of $N$.

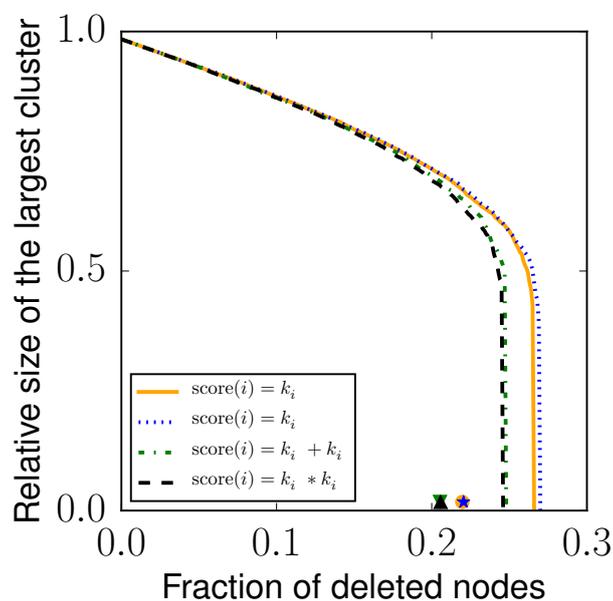

Figure S-2.   Performance of the HDA method on the same multiplex network considered in Figure S-1. Four different plots correspond to four different type of defining the degree of a node.

Like the HD case, we can define at least four methods to define the degree of a node. Please notice that in the multiplex version, when we update the degrees, we exclude those neighbors that are not in the Giant Mutually Connected Component (GMCC) of the network. Figure S-2 shows the effectiveness of the HDA algorithm for the different definitions of nodes' degrees. Similar to the results for HD, it is more effective to combine the scores of different layers, than considering layers as isolated networks. In all the subsequent sections and in the main script, when we refer to HDA, we mean the one in which the score of a node is defined as the multiplication of its degrees across all the layers.

### C. Collective Influence (CI)

In the monoplex version of the CI algorithm [S1], the score $\mathrm{CI}_i(l)$ of node $i$ is equal to the excess degree of $i$ multiplied by the sum of the excess degrees of its neighbours at a specific distance $l$ from $i$:

$$\mathrm{CI}_i(l) = (k_i - 1) \sum_{j \in \partial \mathrm{Ball}(i,l)} (k_j - 1),\tag{S.1}$$

where $\partial \mathrm{Ball}(i,l)$ denotes the neighbors of $i$ at the distance $l$ (i.e., the nodes that have a geodesic distance $l$ from $i$). At each step, the CI score is adaptively calculated for all the

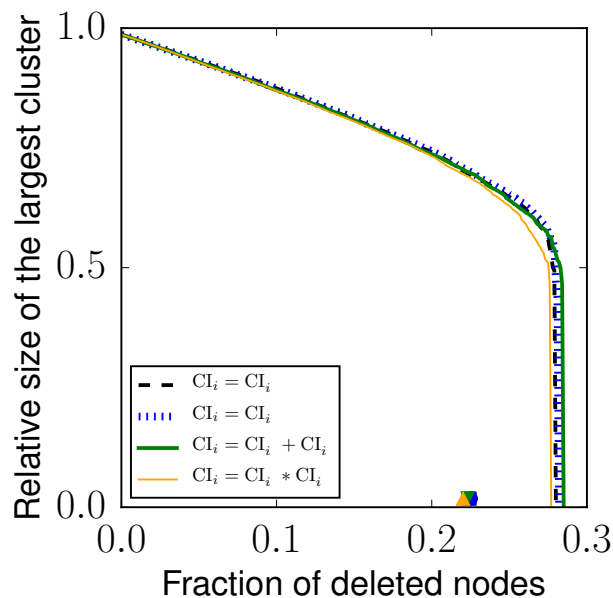

Figure S-3. The performance of the CI algorithm with $l = 4$ on the network of Figure S-1 using different definitions for the collective influence of a node.

nodes; then nodes with the highest score are removed from the network, until the network is dismantled. It was shown [S1, S2] that the performance of the CI method increases with $l$ up to $l = 4$; for $l > 4$ the performance is not improved appreciably as $l$ is increased.

To adapt the CI algorithm to multiplex networks with two layers, we considered several possible definitions of the CI score in the multiplex: (i) using the CI obtained based only on the structure of layer A, (ii) based only on the structure of layer B, (iii) the sum of the CIs of a node in layer A and layer B, and (iv) the product of these two CI scores. Figure S-3 illustrates that, the generalizations of the CI method we considered here are not as effective as those derived based on the HD (Figure S-1) and HDA (Figure S-2) methods. Thus, methods based on the CI measures of the layers do not provide an effective algorithm for the optimal percolation problem.

### D.  Explosive Immunization (EI)

The EI algorithm is based on a method referred to as explosive percolation. The original explosive percolation method was introduced by Achlioptas *et al.* [S3]. In this method at first all the edges are removed; then they are gradually reintroduced to the network, but in a specific order that prevents the formation of the GCC, until a point where the formation of the GCC is inevitable. To add a new edge, first several random edges are selected. Then a score is calculated for each of the selected edges using a predefined kernel[2]. Then the edge with the minimum score is added back the network. The scores represent the contribution of each edge in the formation of the giant cluster. When the network reaches the point where the formation of the giant cluster is inevitable, the rest of the edges are added back using the same kernel.

A problem related to explosive percolation is the optimal immunization [S4] in which the goal is to find the blocker nodes, which if get vaccinated, the giant connected component of the susceptible nodes breaks down; this break down eliminates a large scale epidemic spread. Clusella *et al.* [S4] proposed a reverse approach to find the blockers. They introduced an algorithm that locates instead all the nodes that are irrelevant to the formation of the giant susceptible cluster. In this respect, their algorithm is a modified version of the explosive per-

---

[2] A possible kernel, for example, defines the score as the sum of the sizes of the two clusters connected by the corresponding edge.

colation. Their algorithm considers the site percolation version of the explosive percolation, in which all the links are present but, in the beginning, all the nodes are absent. Then, all the non-blocker nodes (that have no contribution to the formation of the giant susceptible cluster) are added gradually. The remaining nodes are the blocker nodes which should be vaccinated. We refer to this method as the explosive immunization (EI) algorithm.

For a monoplex network we implement the EI algorithm as follows. At each step, we select $N^{(C)} = 1000$ candidate nodes from the set of absent nodes, and calculate the score $\sigma_i$ of each them using the following kernel:

$$\sigma_i = \sum_{j \subset N_i} \left( \sqrt{|C_j|} - 1 \right) + k_i^{(\text{eff})}, \tag{S.2}$$

where, $N_i$ represents the set of all connected components (CCs) linked to node $i$, each of which has a size $C_j$, and $k_i^{(\text{eff})}$ is an effective degree attributed to each node (please see Ref. [S4] for the details). Then the nodes with the lowest scores are added to the network. This procedure is continued until the size of the GCC exceeds a predefined threshold $g^*$ (For the simulations of this paper we used $g^* = \sqrt{N}$). The minus one term in Eq. (S.2) is excluding any leaves connected to node $i$, since they do not contribute to the formation of

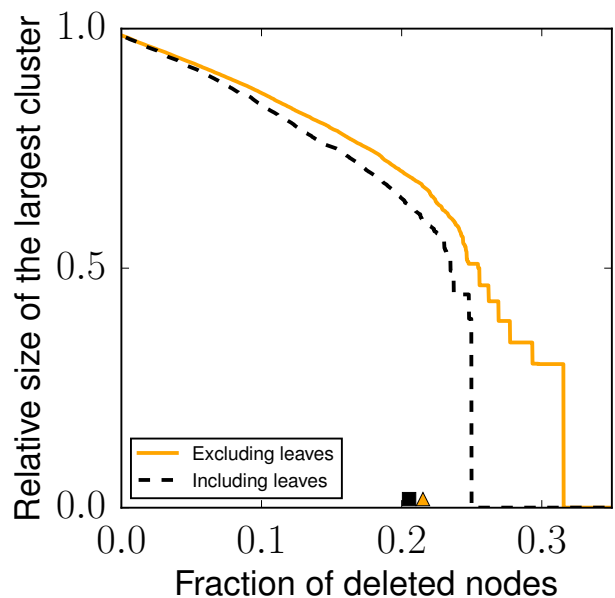

Figure S-4. Comparison between the performance of the EI method when we use Eq. (S.3) that does not exclude the leaves and when we exclude the effect of leaves by replacing $\sqrt{|M|}$ in Eq. (S.3) with $\sqrt{|M|} - 1$. The results correspond to the same network as in Figure S-1.

the GCC and should be ignored in the score of a node.

In our extension of the EI method to multiplex networks, we consider the different kernel but otherwise perform the exact same procedure as the one described above. The new kernel (Eq. (S.3)) we use is based on the sizes of the mutually connected components (MCCs) rather than on the sizes of CCs:

$$\sigma_i = 1/2 \left[ \sum_{j \subset N_i^{[\mathrm{A}]}} \left( \sqrt{|M_j|} \right) + \sum_{j \subset N_i^{[\mathrm{B}]}} \left( \sqrt{|M_j|} \right) \right] + \sqrt{k_i^{[\mathrm{A}](\mathrm{eff})} k_i^{[\mathrm{B}](\mathrm{eff})}}, \tag{S.3}$$

where $N_i^{[\mathrm{A}]}$ is the set of neighbors of node $i$ in layer A, $M_j$ is the size of the MCC to which node $j$ belongs, and $k_i^{[\mathrm{A}](\mathrm{eff})}$ is the effective degree of $i$ in layer A obtained using the same definition proposed [S4] for the monoplex version of the EI method.

In Eq. (S.3) we do not add a minus 1 term to exclude the leaves; the reason is that while in monoplex networks leaves do not have a significant contribution in the formation of the GCC, in multiplex networks even a leaf node is important in the formation of the GMCC. This is because at the sub-critical regime of multiplex networks usually most of the MCCs are isolated nodes or have very small sizes. Figure S-4 certifies that if the leaves were excluded instead, the performance of the algorithm would decrease. Figure S-5 shows that the performance of the EI method does not depend appreciably on the number of candidate

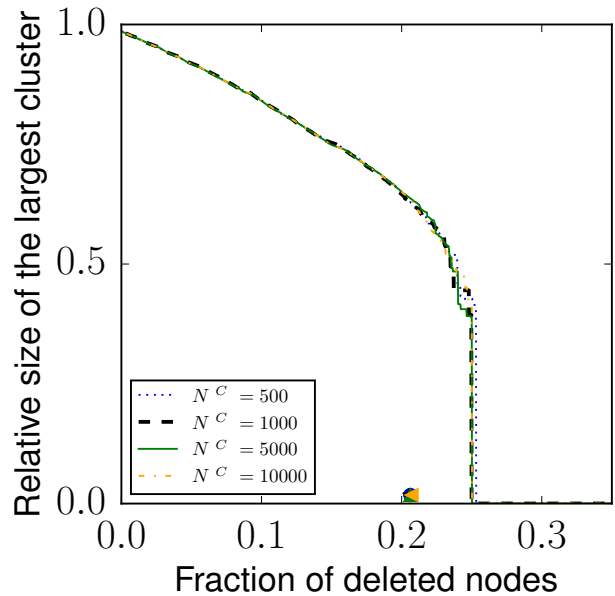

Figure S-5. The results of the EI method do not depend appreciably on the number of candidate nodes $N^{(\mathrm{C})}$. The simulations are performed on the same network used in Figure S-1.

nodes ($N^{(\mathrm{C})}$). In the simulations of Figure 1 of the main text, we used a $N^{(\mathrm{C})} = 1000$.

## E. Simulated Annealing (SA)

The simulated annealing (SA) method has been used for the dismantling problem in monoplex networks [S5]. Generally an SA algorithm defines an energy function that attributes energy values to each configuration of the system. The phase space of the system is searched for the optimal configuration (the one with the minimum energy) by Markov Chain Monte Carlo moves that switch the system from one configuration to another. In dismantling of multiplex networks, the algorithm should find the minimal set of nodes which if deleted the size of the GMCC becomes non-extensive. Each configuration of the multiplex network is represented by $\{R, g\}$, where $R$ and $g$ are, respectively, the number of removed nodes (each node and all its corresponding replica nodes are counted as one node), and the relative size of the GMCC. The energy of a configuration is defined as follows:

$$\varepsilon = Rv + g, \tag{S.4}$$

where $v$ is the cost of removing a node from the multiplex network and in the simulations presented in this paper it is set $v = 0.6$. At each step $t$ of the algorithm, one node, present or removed, is selected at random; then one of the following sets of operations are performed:

- If the node is present and it belongs to the GMCC, it is removed (thus $R_t = R_{t-1} + 1$) and the new size of the GMCC ($g_t$) is calculated.

- If the node is present but it does not belong to the GMCC, then it is removed (thus $R_t = R_{t-1} + 1$); but since it did not belong to the GMCC, $g_t = g_{t-1}$.

- If the node is in the set of removed nodes, it is added back to the network and $R_t = R_{t-1} - 1$. Then $M_i$ (the size of the MCC formed after inserting $i$) is calculated and $g_t = \max(M_i, g_{t-1})$.

Afterwards the energy of the new configuration $\varepsilon_t$ is calculated and the set of operations is accepted with a probability equal to $\min\left(1, \mathrm{e}^{-\beta(\varepsilon_{\mathrm{new}} - \varepsilon)}\right)$. If it is accepted, the new configuration ($\{R_t, g_t\}$) is retained, otherwise, the operations are omitted and the old configuration ($\{R_{t-1}, g_{t-1}\}$) is preserved.

Here, $\beta$ is interpreted as the inverse of the temperature of the annealing process. The SA algorithm starts with a $\beta_{\min}$ and, at each step, $\beta$ is slightly increased by $\delta\beta$. A smaller $\delta\beta$

means a slower decrease in the temperature which allows the SA method to better search for the optimal configurations, at the expense of increasing the running time of the algorithm. In this paper we change the values of $\beta$ from 0.5 to 20.0 with $\delta\beta = 10^{-6}$.

In Figure 1 of the main text, we show that the SA method outperforms the four score-based algorithms; thus, for the analysis of the optimal percolation problem, we mostly use the SA method (see the main text). In Sec. S.3, we also provide results for the second best algorithm, i.e., the HDA method and show that the results are qualitatively similar to those of the SA method.

## S.2. GREEDY REINSERTING (GR)

After a network (either isolated or multiplex) is dismantled using a greedy or score-based algorithm, there are some removed nodes that if added back to the network, the size of the GMCC is not increased substantially, i.e., no cluster with an extensive size is created if they are reinserted. Such nodes may have been removed because the greedy or score-based algorithms are not exact in the sense that they do not take into account the collective nature of the dismantling problem. An approach that addresses this issue is referred to as the greedy reinserting (GR) method [S2, S5]. In the GR method, after a set of structural nodes is detected using another algorithm, at each step a randomly chosen node from the set is reinserted to the network, and unless its reinsertion does not increase the size of the GMCC to a threshold $\sqrt{N}$, it is removed again. This process is continued until practically none of the nodes remained in the set can be added to the network without keeping the size of the GMCC non-extensive.

As shown in Figures S-1–S-5 and Figure 1 of the main text, the GR method boosts effectively the performance of every one of the score-based dismantling algorithms and returns sets of structural nodes with almost identical sizes irrespective of the initial algorithm used. Moreover, the result of each of the score-based algorithms combined with the GR method is nearly as good as that of the SA method (the SA method itself is not improved appreciably by applying a GR method afterwards). These results suggest that probably the sets obtained by the SA method and any one of the score-based algorithms combined with GR are to a considerable extent similar to each other. In Sec. S.3 we show that actually the results of HDA and those of SA are qualitatively similar.

## S.3. COMPLEMENTARY RESULTS FOR THE DEGREE-BASED METHODS

In this section, we provide further results for the HD (Figure S-6) and the HDA (Figure S-7) methods, and also for the combination of the GR method with HDA (Figures S-8–S-9); we compare these results with some of the results of the SA algorithm presented in the main text. In contrast to the SA algorithm, the degree-based algorithms are much more efficient in terms of the running time; hence, we were able to produce some of the results (see Figures S-6–S-7) for larger ER networks.

Figures S-6 and S-7 show that, for both HD and HDA performed on the aggregated representation, the behavior of $q_c$ (the relative size of the set of structural nodes) with

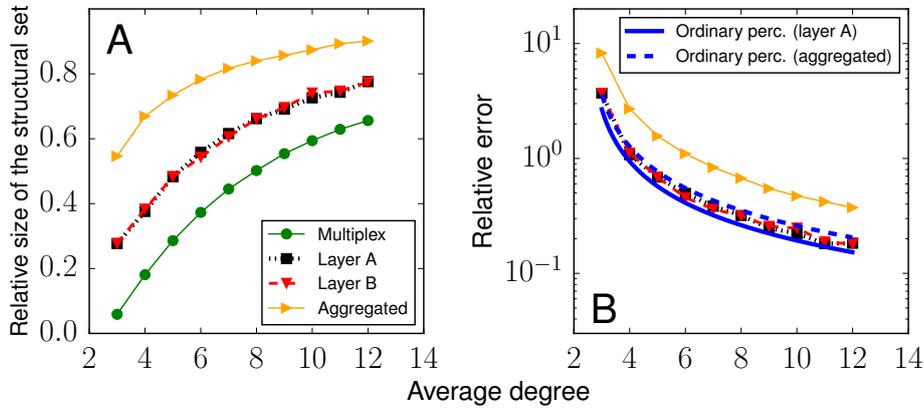

Figure S-6. Like Figure 2 of the main script. Dismantling of a multiplex network with each layer generated independently according to the Erdös-Rényi model with average degree $\langle k \rangle$ and $N = 10^5$ using the HD method.

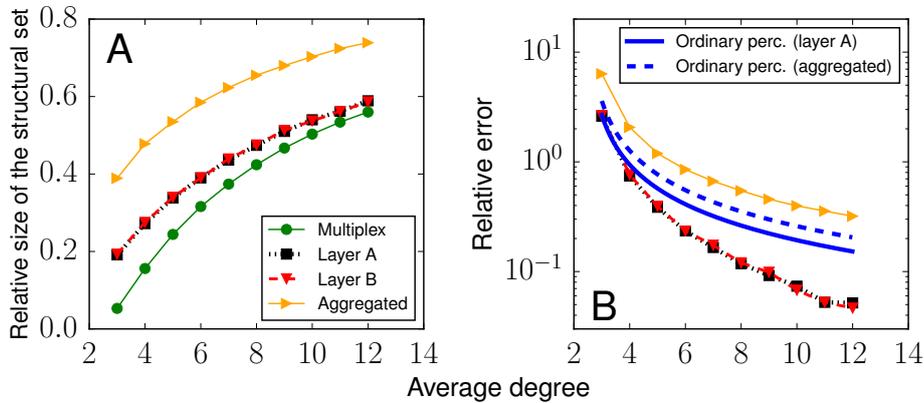

Figure S-7. Like Figure S-6, for the HDA algorithm.

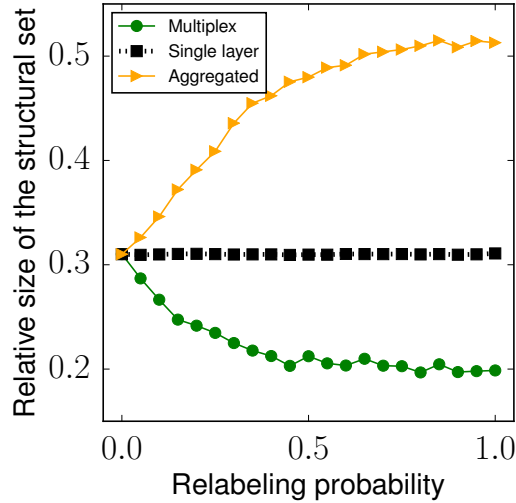

Figure S-8. Optimal percolation results the using HDA+GR method on the networks of Figure 3 of the main script.

respect to the network average degree resembles the results of the SA algorithm. On the other hand, HDA+GR matches better to the result of SA for optimal percolation on each of the single layers of the multiplex network. In particular, for networks with sufficiently large degree, HDA on each of the layers can find a $q_c$ very close to the $q_c$ it obtains for the multiplex representation.

As shown in Figure S-8, the behavior of $q_c$ with respect to the relabeling probability[3] obtained with the HDA+GR method is qualitatively similar to the results of the SA algorithm (Figure 3 of the main text). Moreover, Figure S-9A shows that, as expected, HDA+GR has a relatively higher self-consistency compared to that of the SA algorithm reported in Figure 4 of the main text. It is worth noting that, in contrast to SA, the self-consistency of HDA+GR does not decrease with the relabeling probability in the multiplex representation (Figure S-9A).

Interestingly, despite the qualitative similarity of the results, HDA+GR returns higher values of precision (Figure S-9B), recall (Figure S-9C), and $F_1$-score (Figure S-9D) than those of SA (see Figure 4 of the main text). As there is not much randomness in HDA+GR, the structural nodes are dominantly determined by the sequence of (adaptive) degrees of the nodes and the sets from different network representations have a higher overlap compared

---

[3] higher relabeling probability indicates lower density of overlapping edges and lower interlayer degree-degree correlation.

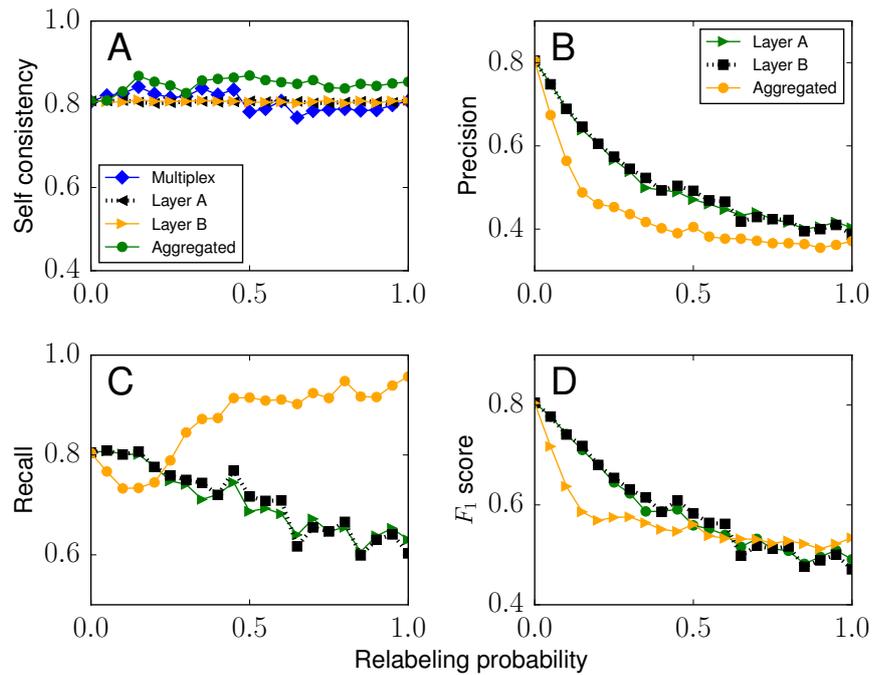

Figure S-9. Results for the performance of the HDA+GR method using the same measures employed in Figure 4 of the main script.

to those find by the SA algorithm.



\end{document}